\documentclass[review,12pt,numbers]{elsarticle}
\usepackage{lineno,hyperref}
\usepackage{amsmath,amssymb}
\usepackage{graphicx}
\usepackage{booktabs}
\usepackage{multirow}
\usepackage{xcolor}
\usepackage{url}
\usepackage{placeins}
\modulolinenumbers[5]
\journal{Computers \& Security}
\begin{document}
\begin{frontmatter}
\title{Quantum ROP: Using Quantum Algorithms for ROP
      Chain Selection in Exploit Construction}
\author[comp]{Carlos Benitez}
\ead{carlos@platinumciber.com}
\address[comp]{PLATINUM CIBER, Argentina}
\begin{abstract}
The quantum computing threat to cybersecurity is nowadays predominantly
framed around Shor's algorithm and its eventual capacity to break
asymmetric cryptography. Beyond cryptanalysis, however, quantum
computing may also enable other capabilities in offensive security. This work
explores one such direction: the application of quantum combinatorial
optimization to Return-Oriented Programming (ROP) gadget selection for exploit
construction.

We formulate gadget selection as a Quadratic Unconstrained Binary
Optimization (QUBO) problem that captures individual gadget cost and
inter-gadget register-clobbering interactions, and solve it using QAOA on
real IBM Heron r2 hardware. Applied to a Linux kernel exploitation scenario,
the QAOA-selected chain achieves privilege escalation to \texttt{uid=0} with
SMEP and SMAP active. Across eight Linux binaries and 16 benchmark
instances, QAOA recovered the lowest-cost valid chain in 11 cases; in the
remaining five, it did not recover the optimum, with the failures associated
with excessive circuit depth on current limited hardware.
\end{abstract}


\end{frontmatter}
%
%


\section{Introduction}
\label{sec:intro}

The dominant association between quantum computing and cybersecurity
nowadays is Shor's algorithm~\cite{shor1994}: a polynomial-time
factoring method that would break asymmetric cryptography once
fault-tolerant hardware with thousands of logical qubits becomes
available~\cite{nist2024pqc,gidney2021factor}.
This framing is well-founded, but it leaves unexplored a broader
question: what can quantum computers already do in offensive security?

This paper answers that question for one specific problem:
the selection of Return-Oriented Programming (ROP) gadget chains
for exploit construction. We formulate gadget selection as a
Quadratic Unconstrained Binary Optimization (QUBO) problem,
solve it using QAOA, and execute it on real IBM Heron r2
quantum hardware. On current NISQ devices, the algorithm
produces valid, ranked exploit chains; the main limitation
is that circuit depth constrains the problem sizes the hardware
can handle reliably. As qubit counts grow and error rates
decrease, these constraints will relax, enabling significantly
larger and more complex gadget selection instances than those
demonstrated here.

\subsection{This Work}
ROP repurposes short instruction sequences ending in \texttt{ret}---\emph{gadgets}---in
the target binary, enabling arbitrary computation without injecting
new code~\cite{shacham2007geometry,roemer2012rop}. Constructing a valid
chain requires that no gadget clobbers a register needed by a later
member, subject to constraints that grow combinatorially in large
binaries. Existing tools (ROPgadget~\cite{ropgadget},
angrop~\cite{angrop}) rely on heuristic search; none formulates
gadget selection as a QUBO suitable for QAOA.

We formulate ROP chain selection as a QUBO encoding individual
gadget cost and inter-gadget register clobbering, and benchmark it
against ES, greedy, and SA on eight Linux binaries. To maintain
full experimental control, we validate the end-to-end pipeline on
the hxpCTF 2020 \texttt{kernel-rop} challenge~\cite{hxpctf2020}:
14,707 gadgets from \texttt{vmlinux} 5.9.0-rc6+ reduced to a
10-qubit QUBO, executed on IBM Heron r2, producing a functional
\texttt{commit\_creds} chain achieving \texttt{uid=0} with SMEP
and SMAP active. Our contribution is demonstrating that a quantum
processor can participate in a functional exploit pipeline.

\subsection{Contributions}

\begin{enumerate}
\item We formulate ROP chain selection as a QUBO problem that captures
      both individual gadget cost and inter-gadget clobbering
      interactions, and demonstrate its execution on real quantum
      hardware.
\item We present a complete, reproducible pipeline from binary
      analysis to exploit validation: \texttt{ROPgadget} $\to$
      \texttt{analyze\_binary} $\to$ \texttt{verify\_gadgets}
      $\to$ QAOA on IBM Heron r2 $\to$ exploit execution.
\item We apply this pipeline to the hxpCTF 2020 \texttt{kernel-rop}
      challenge, producing a functional \texttt{commit\_creds}
      privilege-escalation chain that achieves \texttt{uid=0}
      on Linux 5.9.0-rc6+ with SMEP and SMAP active.
\item We validate the gadget selection process through a
      control experiment: selecting a higher-cost chain with a
      dirty gadget that corrupts the stack layout produces a
      kernel panic, demonstrating that stack-delta constraints
      absent from the QUBO can determine exploit outcome.
\item We empirically characterize the NISQ applicability conditions
      for this problem class and identify circuit depth as
      a contributing factor, with an empirical association between circuit depth
      and valid solution rate on IBM Heron r2.
\item We establish quantum-assisted exploit construction as a
      near-term research direction in offensive security tooling.
\end{enumerate}

\section{Background}
\label{sec:background}

\subsection{Return-Oriented Programming and Kernel Exploitation}

Return-Oriented Programming (ROP) is a code-reuse attack technique
developed to bypass non-executable memory protections such as
$W \oplus X$~\cite{roemer2012rop}. The attacker identifies short
instruction sequences ending in a \texttt{ret} instruction, known
as \emph{gadgets}, in the target binary and chains them by
controlling the stack. Each gadget executes its instructions and
transfers control to the next gadget address placed on the stack,
enabling arbitrary computation without injecting executable code.

In Linux kernel exploitation, hardware mitigations such as SMEP
(Supervisor Mode Execution Prevention) and SMAP (Supervisor Mode
Access Prevention) prevent the kernel from executing or accessing
userspace memory, eliminating shellcode injection and ret2usr
techniques. The canonical kernel LPE chain operates entirely within
kernel text, invoking \texttt{commit\_creds(\&init\_cred)} to assign
root credentials to the calling process, followed by a clean return
to userspace via
\texttt{swapgs\_restore\_regs\_and\_return\_to\_usermode}.
Notably, the \texttt{syscall} instruction executed from kernel mode
transfers control to the kernel's own syscall entry handler
(\texttt{entry\_SYSCALL\_64}), making it unsuitable as a chain
terminator in this context.

Constructing a functional chain requires finding gadgets that load
each required register, ensuring no gadget clobbers a register
needed by a later chain member, and minimizing execution overhead
to maximize reliability. The combinatorial complexity of gadget
selection grows with the number of candidates per role and the
density of inter-gadget register dependencies.

\subsection{QUBO, QAOA, and NISQ Hardware}

A Quadratic Unconstrained Binary Optimization (QUBO) problem is
defined over binary variables $\mathbf{x} \in \{0,1\}^n$:
\begin{equation}
  \min_{\mathbf{x} \in \{0,1\}^n}\; \mathbf{x}^\top Q\, \mathbf{x}
  \label{eq:qubo}
\end{equation}
where $Q \in \mathbb{R}^{n \times n}$ encodes both linear
(diagonal) and quadratic (off-diagonal) terms. The diagonal entry
$Q_{ii}$ represents the individual cost of selecting gadget $i$;
the off-diagonal entry $Q_{ij}$ encodes the penalty for selecting
gadgets $i$ and $j$ together, either because they occupy the same
role (one-hot constraint) or because one clobbers a register
required by the other (cross-penalty). QUBO is NP-hard in general
and subsumes a broad class of combinatorial optimization
problems~\cite{kochenberger2014qubo}. Standard penalty-based
techniques for encoding constraints into QUBO are well
established~\cite{glover2019tutorial}.

QAOA was introduced by Farhi, Goldstone, and
Gutmann~\cite{farhi2014qaoa} as a hybrid quantum-classical
variational algorithm. Given a cost Hamiltonian $H_C$ derived from
the QUBO matrix and a mixer Hamiltonian $H_B$, QAOA prepares the
parametrized state:
\begin{equation}
  |\psi(\boldsymbol{\gamma},\boldsymbol{\beta})\rangle
  = \prod_{k=1}^{p}
    e^{-i\beta_k H_B}\,e^{-i\gamma_k H_C}\,
    |{+}\rangle^{\otimes n}
  \label{eq:qaoa}
\end{equation}
where $p$ is the number of alternating layers,
$\boldsymbol{\gamma} = (\gamma_1,\ldots,\gamma_p)$ are the
phase angles applied to the cost Hamiltonian $H_C$, and
$\boldsymbol{\beta} = (\beta_1,\ldots,\beta_p)$ are the
mixing angles applied to the transverse-field mixer $H_B$;
both are classical parameters optimized to minimize
$\langle H_C \rangle = \langle\psi|H_C|\psi\rangle$.
Each additional layer increases circuit expressibility at the
cost of proportionally greater depth. Parameters are tuned
using SPSA (Simultaneous Perturbation Stochastic
Approximation)~\cite{spall1992spsa}, a gradient-free optimizer
requiring only two circuit evaluations per iteration regardless
of parameter count.

The NISQ era~\cite{preskill2018nisq} is characterized by processors
with tens to thousands of physical qubits operating without full
error correction. For IBM Heron superconducting processors,
two-qubit gate error rates of $\approx 0.1$--$0.3$\% per operation
limit the effective circuit depth before decoherence dominates the
output distribution. The transpiled circuit depth grows with both
the number of qubits and the density of quadratic terms in the
QUBO matrix. We empirically characterize the resulting applicability
window in Section~\ref{sec:results}.

\section{QUBO Formulation for ROP Chain Selection}
\label{sec:formulation}

\subsection{Problem Definition}

Let $\mathcal{R} = \{r_1,\ldots,r_m\}$ be the set of roles required
by the target ROP chain. For each role $r_j$, automated disassembly
and semantic verification (Section~\ref{sec:pipeline}) produce a
pool of verified candidate gadgets
$\mathcal{G}_j = \{g_{j,1},\ldots,g_{j,n_j}\}$.
Each gadget $g_{j,k}$ carries an individual cost $c_{j,k} \in \mathbb{Z}^{+}$
and a clobber set $\mathcal{C}(g_{j,k})$ of registers modified
as side effects beyond the intended register load:
\begin{equation}
  c(g) = |\mathcal{C}(g)| \times 3 + (n_{\mathrm{insns}}(g) - 1)
  \label{eq:gadgetcost}
\end{equation}
where $n_{\mathrm{insns}}(g)$ is the instruction count up to and
including the terminal \texttt{ret}. The factor of 3 weights each
clobbered register more heavily than an extra instruction, reflecting
the functional impact of silent register corruption on subsequent
chain members. A clean \texttt{pop~rdi;~ret} gadget has cost 1;
a gadget that also clobbers \texttt{rsi} carries cost 4.

We define a binary conflict indicator
$\delta(g_{j,k}, g_{l,m}) \in \{0,1\}$
that equals 1 when a clobber in gadget $g_{j,k}$ corrupts
a register required by a later role $r_l$ ($j < l$):
\begin{equation}
  \delta(g_{j,k}, g_{l,m}) =
  \begin{cases}
    1 & \text{if } \exists\, r \in \mathcal{C}(g_{j,k})
          \text{ s.t.\ } r \in \mathrm{needs}(r_l),\; j < l \\
    0 & \text{otherwise}
  \end{cases}
  \label{eq:cross}
\end{equation}
where $\mathrm{needs}(r_l)$ is the set of registers that role $r_l$
and all subsequent roles must receive unmodified.
For the \texttt{commit\_creds} chain, $\mathrm{needs}(\texttt{load\_rsi})$
includes \texttt{rdi}, $\mathrm{needs}(\texttt{load\_rdx})$
includes \texttt{rdi} and \texttt{rsi}, and so on.

We introduce a binary variable $x_{j,k} \in \{0,1\}$ for each
candidate gadget, where $x_{j,k} = 1$ indicates selection of
$g_{j,k}$ for role $r_j$. The total variable count
$n = \sum_j n_j$ equals the number of qubits required.
With $m = 5$ roles and $n_j = 2$ candidates per role, the
\texttt{commit\_creds} instance requires $n = 10$ qubits and
covers a search space of $2^{10} = 1024$ binary assignments,
of which exactly $2^5 = 32$ satisfy the one-hot constraint.

\subsection{QUBO Matrix Construction}

The objective function combines three terms:
\begin{align}
  E_{\mathrm{cost}} &=
    \sum_j \sum_k c_{j,k}\, x_{j,k}
    \label{eq:ecost}\\
  E_{\mathrm{one\text{-}hot}} &=
    A_{\mathrm{pen}}\sum_j
    \Bigl(1 - \sum_k x_{j,k}\Bigr)^2
    \label{eq:eonehot}\\
  E_{\mathrm{cross}} &=
    A_{\mathrm{cross}}\sum_{j < l}\sum_k\sum_{m}
    \delta(g_{j,k},g_{l,m})\, x_{j,k}\, x_{l,m}
    \label{eq:ecross}
\end{align}
\begin{equation}
  E(\mathbf{x}) =
    E_{\mathrm{cost}} +
    E_{\mathrm{one\text{-}hot}} +
    E_{\mathrm{cross}}
  \label{eq:obj}
\end{equation}

$E_{\mathrm{cost}}$ encodes the sum of individual gadget costs
for the selected gadgets. $E_{\mathrm{one\text{-}hot}}$ enforces
exactly one gadget per role: the penalty coefficient
$A_{\mathrm{pen}}$ must be large enough that no one-hot
violation reduces the total energy below any valid solution.
For QAOA we use a fixed $A_{\mathrm{pen}} = 10$, calibrated
for our benchmark instances; for Simulated Annealing we set
$A_{\mathrm{pen}}$ dynamically as described in
Section~\ref{sec:greedy}. $E_{\mathrm{cross}}$ assigns penalty $A_{\mathrm{cross}}$
to each conflicting pair from different roles. The constraint
$j < l$ avoids double-counting, so each conflicting pair
receives exactly one penalty of $A_{\mathrm{cross}} = 8$.

These terms assemble into the QUBO matrix $Q$ as follows.
The diagonal entry $Q_{ii}$ receives the individual gadget cost
$c_{j,k}$ for the variable $x_{j,k}$ mapped to index $i$,
offset by the one-hot expansion of~\eqref{eq:eonehot}.
We use the upper-triangular QUBO convention, so each
off-diagonal coefficient appears exactly once in the objective.
An off-diagonal entry $Q_{ij}$ ($i < j$) is set to
$2A_{\mathrm{pen}}$ when variables $i$ and $j$ share the same
role, and to $A_{\mathrm{cross}}$ when they belong to different
roles with a register interaction. All other off-diagonal entries
are zero. In our 10-qubit instance, 17 of the 45 possible
off-diagonal pairs carry non-zero values. The matrix $Q$ is converted to an
Ising Hamiltonian via $x_i = (1 - \sigma_i^z)/2$, yielding a
cost Hamiltonian $H_C$ decomposed into Pauli $Z$ and $ZZ$
operators for circuit construction.

\subsection{Penalty Calibration and Greedy Baseline}
\label{sec:greedy}

For Simulated Annealing, the penalty coefficient must be set
dynamically to guarantee feasibility across instances with
varying maximum gadget cost. We set:
\begin{equation}
  A_{\mathrm{pen}} = \max_{j,k} c_{j,k} + m \cdot A_{\mathrm{cross}} + 1
  \label{eq:penalty}
\end{equation}
which ensures that for SA the energy reduction from violating
a one-hot constraint never compensates the penalty incurred.
QAOA uses a fixed $A_{\mathrm{pen}} = 10$ as noted above.

As a classical reference, a greedy solver selects the gadget of
minimum individual cost independently for each role, without
considering cross-penalties. Greedy runs in $O(\sum_j n_j)$
time and coincides with the optimal solution when all cross-penalties
are zero. When inter-role interactions are dense, greedy may
select a gadget that clobbers a register required by a later role,
yielding a suboptimal chain. In the benchmark instances evaluated
in Section~\ref{sec:results}, all gadget pools contain at least
one clean gadget per role, so greedy matches the ES optimal within the retained candidate
set. The crossover point at which greedy becomes suboptimal
is discussed in Section~\ref{sec:discussion_and_limitations}.

\section{Implementation Pipeline}
\label{sec:pipeline}

\begin{figure}[ht]
  \centering
  \includegraphics[width=\linewidth]{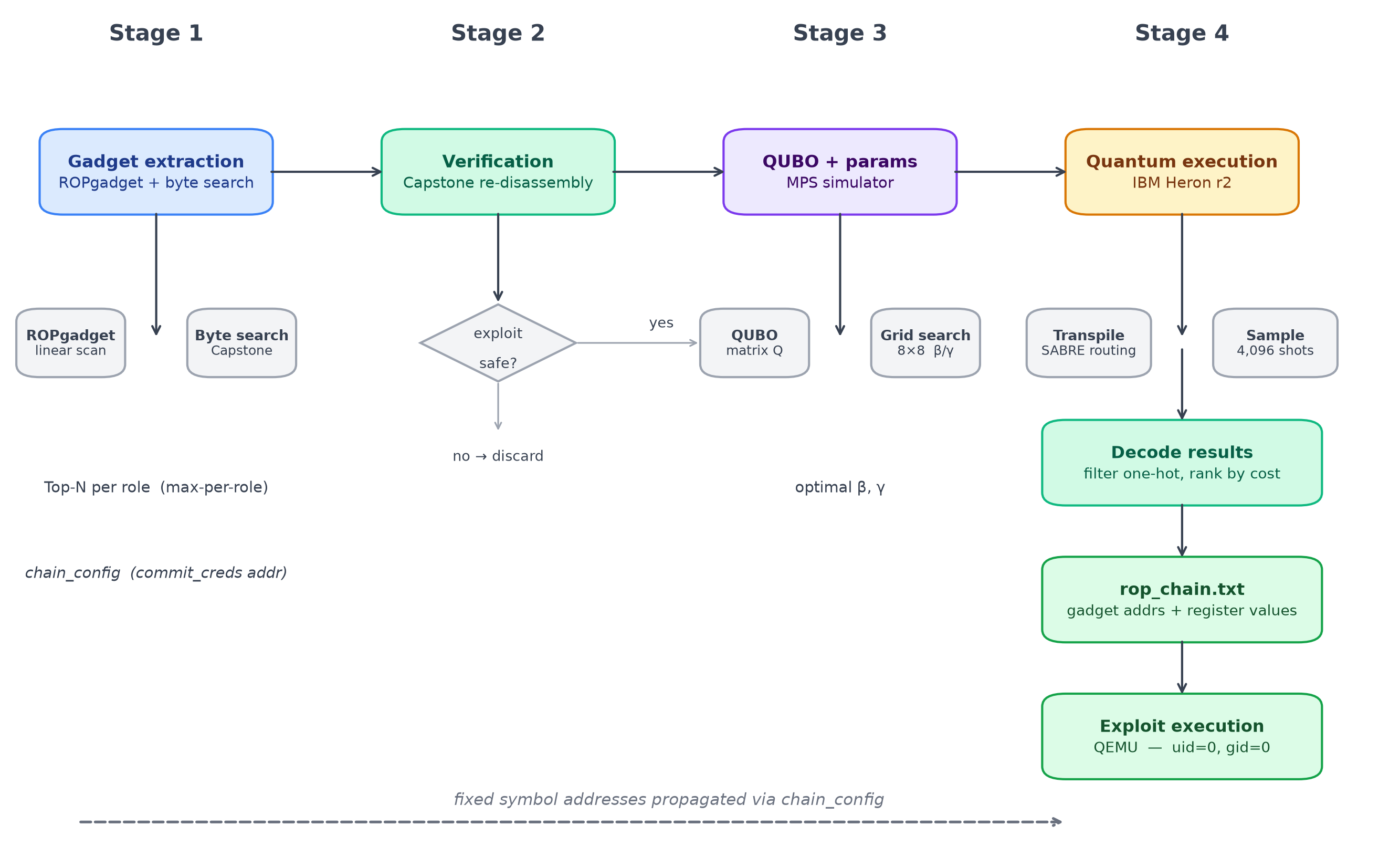}
  \caption{End-to-end pipeline for QAOA-assisted ROP chain selection.
           Gadgets are extracted and verified from the target binary,
           encoded as a QUBO, and optimized on IBM Heron r2 hardware.
           Fixed symbol addresses (\texttt{commit\_creds}, \texttt{init\_cred})
           are propagated via \texttt{chain\_config} without entering the QUBO.}
  \label{fig:pipeline}
\end{figure}

The pipeline transforms a target binary into a QAOA-ready QUBO
instance and delivers a verified ROP chain. It comprises four
sequential stages: gadget extraction, semantic verification,
QUBO construction and parameter optimization, and quantum
execution. Figure~\ref{fig:pipeline} illustrates the full flow.

\subsection{Gadget Extraction and Role Classification}
Raw gadget candidates are extracted using
ROPgadget~\cite{ropgadget} in \texttt{{--}rop} mode over the
full binary image. For a monolithic kernel such as
\texttt{vmlinux}, this typically produces from tens of
thousands to more than one million candidate sequences,
depending on the binary and extraction settings. The output is then filtered and classified
by role using a custom analyzer (\texttt{analyze\_binary}) that
applies two complementary methods.

Candidates are classified using per-role pattern matching on the
ROPgadget output (e.g., any sequence with \texttt{pop rdi} before
\texttt{ret} for \texttt{load\_rdi}) and a Capstone~\cite{capstone}
byte search for gadgets missed by alignment constraints.
The $N$ lowest-cost gadgets per role are retained
(\texttt{--max-per-role}), bounding qubit count to $N\times m$.
Fixed kernel symbol addresses are stored in a
\texttt{chain\_config} key for downstream reconstruction.

\subsection{Semantic Verification}

Each candidate gadget is re-disassembled directly from the
binary ELF using Capstone to confirm that the instruction
sequence matches the intended semantics and terminates at
the first clean \texttt{ret}. This step catches false positives
from pattern-based extraction, including sequences where the
target instruction appears in a data section, sequences with
intervening \texttt{call} or \texttt{jmp} instructions that
would redirect control flow before reaching the \texttt{ret},
and gadgets where the disassembly boundary differs between
ROPgadget and Capstone due to overlapping encodings.

During verification, the clobber set $\mathcal{C}(g)$ is
computed by inspecting all destination registers written
between the first instruction and the \texttt{ret}, excluding
the intended target register. Gadgets are classified as
\emph{exploit-safe} if they can be placed directly in a
kernel-mode ROP chain, or \emph{qaoa-only} if they carry
a clobber that makes them inadvisable for direct deployment.
Only exploit-safe gadgets enter the QUBO. The verified gadget
pool and the \texttt{chain\_config} key are written to a
single JSON file that serves as the interface between
verification and QUBO construction.

\subsection{QUBO Construction and Parameter Optimization}

The verified JSON is consumed by \texttt{qaoa\_binary}, which
constructs the QUBO matrix $Q$ following
Section~\ref{sec:formulation}, maps it to an Ising Hamiltonian
$H_C$ via $x_i = (1 - \sigma_i^z)/2$, decomposes $H_C$ into
Pauli terms, and builds the QAOA ansatz circuit
(Equation~\ref{eq:qaoa}) for the chosen depth $p$.

Variational parameters $(\boldsymbol{\gamma}, \boldsymbol{\beta})$
are found by an $8\times8$ grid search on a local MPS simulator
(4,096 shots per point), followed by optional SPSA refinement
up to 150 iterations. At $n=10$ qubits each simulator evaluation
completes in under one second on commodity hardware.

\subsection{Quantum Execution on IBM Heron r2}

The optimized circuit is transpiled to the native gate set
of the target backend using Qiskit~\cite{qiskit2024} with
optimization level 3 and SABRE routing. Transpilation maps
abstract two-qubit gates to the \texttt{ECR} native gates of
the Heron~r2 architecture and inserts swap networks to satisfy
the device connectivity graph. The transpiled depth and
two-qubit gate count are logged before submission to verify
that the circuit falls within the empirical applicability
window characterized in Section~\ref{sec:results}.

Jobs are submitted to IBM Heron r2 processors via the Qiskit
Runtime Service~\cite{qiskit2024} with 4,096 shots per run.
On completion, \texttt{decode\_ibm\_results} recovers the raw
bitstring distribution, filters for valid solutions satisfying
the one-hot constraint, and ranks them by total gadget cost.
The chain at rank 1 is exported as a \texttt{rop\_chain.txt}
file containing gadget addresses and the values to be placed
on the stack for each register load. The addresses of fixed
symbols (\texttt{commit\_creds} and \texttt{init\_cred}) are
appended from \texttt{chain\_config}, completing the chain.

The valid bitstring rate $V_{\mathrm{rate}}$, defined as the
fraction of distinct measurement outcomes satisfying the
one-hot constraint, is recorded for each job and used as
the primary signal-quality metric in
Section~\ref{sec:results}.

\section{Experimental Evaluation}
\label{sec:results}

\subsection{Experimental Setup}

We evaluated the pipeline across three Linux kernel images and five
userspace binaries, covering both long-term support releases and
actively deployed versions. Table~\ref{tab:binaries} lists the
binaries, versions, and gadget pool sizes after ROPgadget extraction.

\begin{table}[ht]
\centering
\caption{Binary corpus. Pool size is the total number of gadgets
         found by ROPgadget before role classification.}
\label{tab:binaries}
\small
\begin{tabular}{@{}lllr@{}}
\toprule
ID & Binary & Version & Pool size \\
\midrule
K1 & vmlinux & 5.15.0-153 LTS  & 1,122,173 \\
K2 & vmlinux & 6.1.0-060100    & 1,093,575 \\
K3 & vmlinux & 6.8.0-45        & 1,319,889 \\
U1 & sshd    & OpenSSH 8.9p1   &    24,575 \\
U2 & nginx   & 1.18.0          &    47,025 \\
U4$^\dagger$ & bash & 5.1.16   &    59,653 \\
U5 & python3 & 3.10.12         &   233,135 \\
U6 & libc.so.6 & GLIBC 2.35   &   112,047 \\
\bottomrule
\end{tabular}
\begin{minipage}{\linewidth}
\smallskip
\centering
{\small $^{\dagger}$U3 (\texttt{sudo}) was excluded: all five roles
produced empty verified gadget pools. IDs U4--U6 preserve the
original experiment labels.}
\end{minipage}
\end{table}

For each binary, we generated three instance sizes by varying the
maximum gadgets per role: $N=2$ (S-2, $n=10$ qubits),
$N=3$ (S-3, $n=15$ qubits), and $N=4$ (S-4, $n=20$ qubits),
yielding search spaces of 32, 243, and 1,024 combinations
respectively. QAOA-HW experiments were conducted for S-2 and
S-3 only. S-4 instances (depth $>$ 300 after transpilation)
exceed the current NISQ applicability window; they were
evaluated with ES, GR, and SA only and are not reported
in Table~\ref{tab:results}, which covers S-2 and S-3. Three binaries (\texttt{sudo}, \texttt{callme}, \texttt{pivot}) were excluded
because one or more roles produced no verified gadgets after
semantic filtering.

Each instance was solved by four methods: Exhaustive Search (ES),
which enumerates all combinations and reports the optimal
solution within the retained candidate set;
Simulated Annealing (SA), run for 10 independent trials using
the dynamic penalty of Equation~\ref{eq:penalty}; QAOA on a
local MPS simulator (QAOA-SIM), executed as a single run with
the optimal parameters identified by the grid search; and QAOA on IBM Heron r2
hardware (QAOA-HW), run for 5 independent jobs of 4,096 shots
each per instance.
The greedy baseline is computed as the minimum-cost gadget
per role from the verified pool. All corpus benchmark instances
in Sections~\ref{sec:results}--\ref{sec:p} use roles
\texttt{load\_rdi}, \texttt{load\_rsi}, \texttt{load\_rdx},
\texttt{load\_rax}, and \texttt{syscall}, which model the
general register-loading structure common to a broad class
of ROP chains and allow direct comparison across binaries.
The end-to-end case study in Section~\ref{sec:casestudy}
uses the commit\_creds-specific role set described there.

\begin{table}[ht]
\centering
\caption{Optimization results across 16 instances (8 binaries
         $\times$ S-2 and S-3). All cost values represent the
         pure gadget cost of valid solutions satisfying the
         one-hot constraint. ES = optimal within retained
         candidate set. GR = greedy. SA = mean and standard
         deviation over 10 valid solutions; individual-run best
         values were not persisted. QAOA-SIM =
         single run with optimal parameters (pure gadget cost).
         QAOA-HW = best over 5 independent jobs of 4,096 shots
         each (pure gadget cost). A checkmark indicates the
         method found the ES optimal.}
\label{tab:results}
\small
\begin{tabular}{@{}llrrrlrr@{}}
\toprule
Binary & Inst. & ES & GR & SA mean ($\sigma$) &
        QAOA-SIM & HW mean & HW best \\
\midrule
vmlinux 5.15 & S-2 & 22 & 22 & 29.2 (6.0) & 22\checkmark & 36.2 & 22\checkmark \\
             & S-3 & 20 & 20 & 30.8 (7.2) & 20\checkmark & 31.2 & 22 \\
vmlinux 6.1  & S-2 & 22 & 22 & 31.4 (3.6) & 26 & 24.4 & 22\checkmark \\
             & S-3 & 17 & 17 & 33.0 (8.6) & 28 & 42.8 & 33 \\
vmlinux 6.8  & S-2 & 18 & 18 & 26.6 (3.0) & 18\checkmark & 18.0 & 18\checkmark \\
             & S-3 & 17 & 17 & 28.3 (5.8) & 17\checkmark & 19.6 & 17\checkmark \\
sshd 8.9p1   & S-2 & 13 & 13 & 20.4 (3.8) & 20 & 17.2 & 13\checkmark \\
             & S-3 & 13 & 13 & 20.8 (3.9) & 13\checkmark & 25.6 & 20 \\
nginx 1.18   & S-2 & 31 & 31 & 31.0 (0.0) & 31\checkmark & 31.0 & 31\checkmark \\
             & S-3 & 31 & 31 & 31.5 (0.5) & 31\checkmark & 31.6 & 31\checkmark \\
bash 5.1     & S-2 & 25 & 25 & 25.8 (1.0) & 25\checkmark & 25.4 & 25\checkmark \\
             & S-3 & 25 & 25 & 28.4 (2.7) & 25\checkmark & 25.4 & 25\checkmark \\
python3 3.10 & S-2 & 19 & 19 & 26.5 (4.9) & 30 & 26.8 & 24 \\
             & S-3 & 19 & 19 & 27.6 (4.6) & 27 & 30.2 & 28 \\
libc 2.35    & S-2 & 20 & 20 & 24.8 (3.9) & 20\checkmark & 20.0 & 20\checkmark \\
             & S-3 & 20 & 20 & 24.0 (4.0) & 20\checkmark & 21.6 & 20\checkmark \\
\bottomrule
\end{tabular}
\end{table}

\FloatBarrier
\subsection{Optimization Methods Comparison}

Table~\ref{tab:results} reports the ES optimal, greedy
cost (GR), SA mean~($\sigma$) over 10 runs, the result
of a single QAOA-SIM run with optimal parameters, and
QAOA-HW best over 5 independent jobs of 4,096 shots each.
All cost values represent pure gadget cost of valid solutions.

ES finds the optimal within the retained candidate set by construction. The greedy baseline matches the ES optimal in all 16 instances:
the gadget pools for these binaries each contain at least one
clean gadget per role, so selecting the minimum-cost gadget
independently per role coincides with the ES solution within
the retained set. The conditions under which greedy diverges
are discussed in Section~\ref{sec:discussion_and_limitations}.

SA produces valid solutions in every instance; mean and standard
deviation over 10 runs are reported in Table~\ref{tab:results}.

QAOA-SIM with $p=1$ finds the ES optimal in 11 of 16
instances. The five failures (vmlinux 6.1 S-2 and S-3,
sshd S-2, python3 S-2 and S-3) are instances where the
$p=1$ circuit with the retained parameter set does not
recover the ES optimum. For vmlinux 6.1 S-3,
running QAOA with $p=2$ and 100 SPSA iterations recovers
the ES optimal at the cost of a proportionally
deeper circuit.

\begin{figure}[ht]
  \centering
  \includegraphics[width=\linewidth]{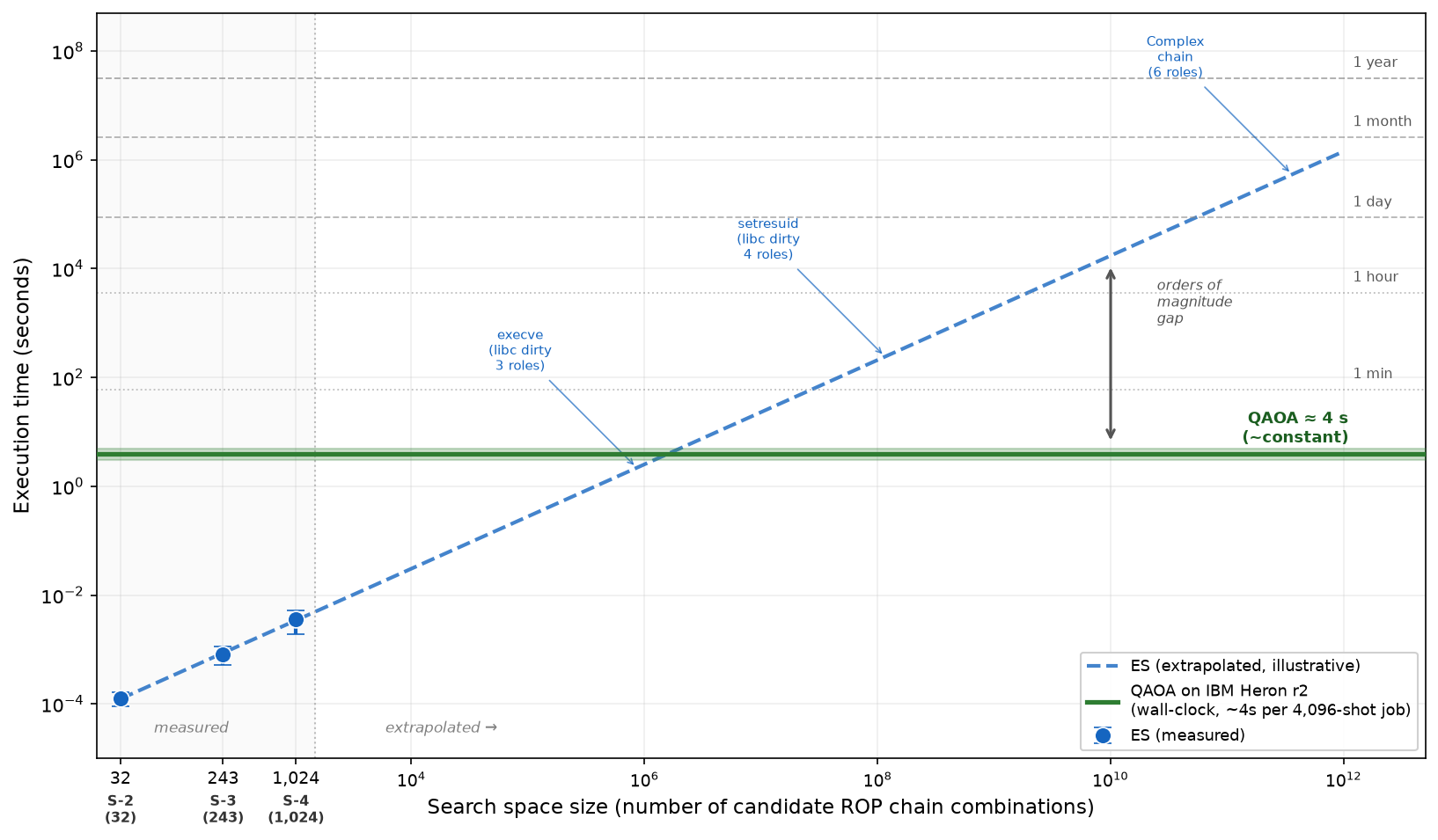}
  \caption{Scalability comparison (illustrative extrapolation).
           Left: ES and SA execution time vs.\ search space size
           (solid points: measured on 8 binaries; dashed lines:
           extrapolated from three instance sizes to $10^{12}$
           combinations). Right: QAOA circuit depth vs.\ number
           of qubits (approximately linear over the measured range). Extrapolations
           are included for illustrative purposes only.}
  \label{fig:scalability}
\end{figure}

QAOA-HW recovers the ES optimum in at least one of five
independent jobs for 11 of 16 instances. The five instances
where no job recovered the optimum all correspond to
transpiled circuit depths exceeding 200. Among the 11 instances with
depth in the lower range, QAOA-HW matches the ES optimal in
every case. Figure~\ref{fig:scalability} shows how the execution time
of ES grows with the number of combinations (illustrative
extrapolation), alongside the QAOA circuit depth as a function
of the number of qubits (approximately linear over the measured
range). These quantities are shown on separate axes and should
not be interpreted as a direct wall-clock comparison; end-to-end
QAOA cost includes parameter optimization, transpilation, and
multiple hardware jobs not reflected in the per-job execution time.

\subsection{NISQ Hardware Characterization}
\begin{figure}[ht]
  \centering
  \includegraphics[width=\linewidth]{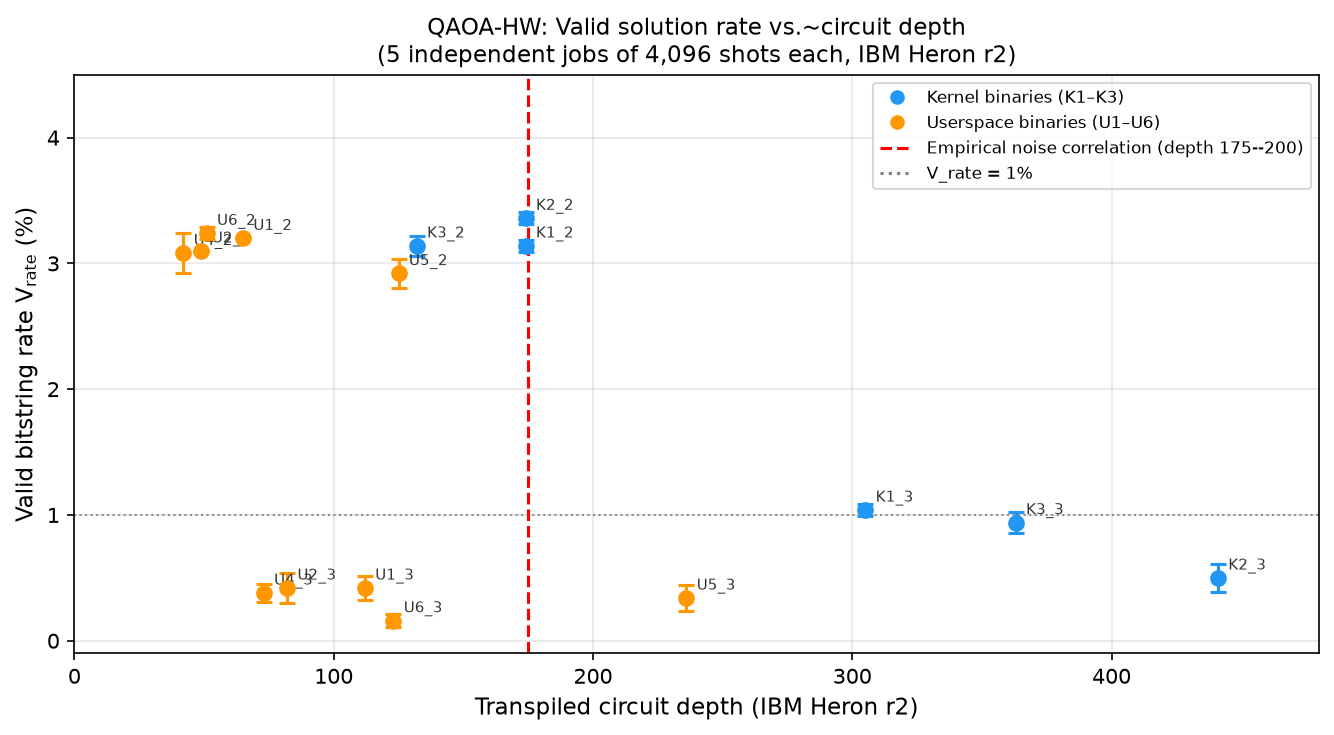}
  \caption{Valid bitstring rate $V_{\mathrm{rate}}$ vs.\ transpiled
           circuit depth on IBM Heron r2 (5 independent jobs of
           4,096 shots each per instance). Blue: kernel binaries.
           Orange: userspace binaries. A marked transition in signal
           quality is observed in the depth range 175--200.}
  \label{fig:vrate}
\end{figure}

Figure~\ref{fig:vrate} plots the valid bitstring rate
$V_{\mathrm{rate}}$ against transpiled circuit depth for all
16 instances across 5 runs on IBM Heron r2 hardware. A
marked transition in signal quality is observed in the
depth range 175--200: instances in the lower portion of
this range show $V_{\mathrm{rate}} > 1\%$ and QAOA-HW
recovers the ES optimal in at least one of the five runs,
while instances above depth 200 show $V_{\mathrm{rate}} < 0.5\%$
and the optimization signal degrades substantially. This
transition should be understood as an empirical correlation
observed on this hardware generation, not as a sharp
architectural threshold.

At approximately 0.2\% two-qubit gate error per ECR gate on
Heron r2, a circuit with transpiled depth 200 (corresponding
to approximately 80--120 two-qubit gates after transpilation)
is associated with a reduction in valid solution yield
(Spearman $\rho = -0.26$, $p = 0.32$, $n = 16$; the
correlation is suggestive but does not reach statistical
significance at this sample size, partly due to the confound
between circuit depth and instance size). As hardware improves and gate error rates
decrease, this correlation may shift; quantifying the extent of
improvement requires hardware-specific noise modeling beyond the
scope of this work.

\subsection{Effect of Circuit Depth $p$}
\label{sec:p}
We compared $p=1$ and $p=2$ on all 16 instances using the
MPS simulator, with SPSA optimizing parameters independently
for each value of $p$ over 100 iterations. With $p=2$, the
optimizer must tune four parameters ($\boldsymbol{\gamma},
\boldsymbol{\beta}$ each of length 2) versus two for $p=1$,
making convergence harder within the same iteration budget.

$p=2$ underperforms $p=1$ in 10 of 16 instances, matches in
4, and improves in 2. The two improvements correspond to
instances with the highest cross-penalty density, where the
additional circuit layer provides expressibility that $p=1$
cannot achieve. For the remaining instances, the optimization
landscape at $p=2$ is sufficiently flat within 100 SPSA
iterations that the optimizer does not reach the same quality
as the well-characterized $p=1$ grid-search optimum. All
hardware experiments reported in this paper use $p=1$.

\subsection{End-to-End Case Study: hxpCTF 2020 kernel-rop}
\label{sec:casestudy}

\subsubsection{Vulnerability and Environment}

The hxpCTF 2020 \texttt{kernel-rop} challenge provides a
controlled kernel exploitation environment based on a stack
buffer overflow in \texttt{hackme\_write}, a function exported
by the \texttt{hackme.ko} kernel module loaded into Linux
5.9.0-rc6+. The overflow allows an attacker to overwrite the
return address of \texttt{hackme\_write} after leaking the
stack canary value via \texttt{hackme\_read}. The kernel is
configured with SMEP and SMAP active and KASLR disabled,
making gadget addresses deterministic and enabling direct
comparison of QAOA-selected chains across runs.

\subsubsection{Gadget Pool and QUBO Instance}

ROPgadget extracts 14,707 candidate sequences from the kernel
image. After role classification with \texttt{analyze\_binary}
and semantic verification with Capstone, the pool reduces to
10 exploit-safe gadgets across a five-role gadget-selection
subproblem, 2 gadgets per role.

The five roles constitute a gadget-selection subproblem
chosen to expose inter-gadget interactions within the exploit
pipeline. They are not the minimal set of instructions
required by \texttt{commit\_creds}: the optimal chain uses
the clean \texttt{pop~rdi;~ret} gadget (cost~1) and therefore
does not strictly require the four auxiliary roles to repair
register state. Their inclusion in the QUBO is motivated by
the presence of dirtier \texttt{load\_rdi} candidates (cost~5)
that clobber \texttt{rsi}, \texttt{rdx}, and \texttt{rcx},
creating 17 cross-penalty terms and providing an instance
with explicit inter-gadget interaction structure.

Table~\ref{tab:gadgets} lists the verified candidates with
their disassembly and costs.

\begin{table}[ht]
\centering
\caption{Verified gadget candidates for the hxpCTF case study.
         Cost computed via Equation~\ref{eq:gadgetcost}.}
\label{tab:gadgets}
\small
\begin{tabular}{@{}lllr@{}}
\toprule
Role & Address & Disassembly & Cost \\
\midrule
load\_rdi & \texttt{0xffffffff81006370} & \texttt{pop rdi; ret}           & 1 \\
          & \texttt{0xffffffff8251251f} & \texttt{pop rdi; pop rsi; ret}  & 5 \\
load\_rsi & \texttt{0xffffffff8150b97e} & \texttt{pop rsi; ret}           & 1 \\
          & \texttt{0xffffffff810e5a9d} & \texttt{pop rsi; pop rdx; ret}  & 4 \\
load\_rdx & \texttt{0xffffffff81007616} & \texttt{pop rdx; ret}           & 1 \\
          & \texttt{0xffffffff8100767b} & \texttt{pop rdx; pop rdi; ret}  & 5 \\
load\_rcx & \texttt{0xffffffff815f4bbc} & \texttt{pop rcx; ret}           & 1 \\
          & \texttt{0xffffffff81a6ee8b} & \texttt{pop rcx; ret}           & 1 \\
xor\_eax  & \texttt{0xffffffff810001ce} & \texttt{xor eax, eax; ret}      & 1 \\
          & \texttt{0xffffffff81003b5b} & \texttt{xor eax, eax; ret}      & 1 \\
\bottomrule
\end{tabular}
\end{table}

The 10 variables produce a QUBO matrix with 17 non-zero
off-diagonal pairs. The ES optimal has cost 5,
selecting the minimum-cost gadget for each role.

\subsubsection{QAOA Execution and Results}

The QAOA circuit for $p=1$ transpiles on IBM Heron r2 to
depth 185 with 80 two-qubit gates. Grid search on the MPS
simulator identifies $\beta = 0.2$, $\gamma = 1.4$ as optimal
parameters. A single job on IBM Heron r2 with 4,096 shots
returns 954 distinct bitstrings, of which 23 satisfy the
one-hot constraint. Table~\ref{tab:top5} reports the five
lowest-cost valid chains from this measurement.

\begin{table}[ht]
\centering
\caption{Top-5 valid ROP chains returned by QAOA-HW for the
         hxpCTF instance. Each row shows the gadget selected
         for each role, its cost, and the number of measurement
         shots in which that configuration appeared.}
\label{tab:top5}
\small
\begin{tabular}{@{}rrrrrrrr@{}}
\toprule
Rank & load\_rdi & load\_rsi & load\_rdx & load\_rcx & xor\_eax & Cost & Shots \\
\midrule
1 & \texttt{...6370} & \texttt{...97e} & \texttt{...616} & \texttt{...8b} & \texttt{...ce} & 1+1+1+1+1=5 & 3 \\
2 & \texttt{...6370} & \texttt{...97e} & \texttt{...616} & \texttt{...8b} & \texttt{...5b} & 1+1+1+1+1=5 & 2 \\
3 & \texttt{...6370} & \texttt{...97e} & \texttt{...616} & \texttt{...bc} & \texttt{...5b} & 1+1+1+1+1=5 & 1 \\
4 & \texttt{...6370} & \texttt{...97e} & \texttt{...616} & \texttt{...bc} & \texttt{...ce} & 1+1+1+1+1=5 & 1 \\
5 & \texttt{...1f}   & \texttt{...97e} & \texttt{...616} & \texttt{...8b} & \texttt{...5b} & 5+1+1+1+1=9 & 5 \\
\bottomrule
\end{tabular}
\end{table}

The four rank-1 through rank-4 configurations all share the
optimal cost of 5 (one clean gadget per role). The rank-5
configuration selects a higher-cost gadget for \texttt{load\_rdi}
and represents the first suboptimal valid chain in the
distribution. Of the 4,096 shots, 7 produce an optimal-cost
chain ($P(\text{optimal}) = 7/4096 \approx 0.17\%$);
$V_{\mathrm{rate}}$, by contrast, measures distinct valid
outcomes as a fraction of distinct bitstrings observed
($23/954 \approx 2.4\%$) and is therefore not directly
comparable to a shot-based success probability.

\subsubsection{Exploit Validation}

The rank-1 chain is exported to \texttt{rop\_chain.txt} with
the fixed \texttt{commit\_creds} address
(\texttt{0xffffffff814c6410}) and \texttt{init\_cred} pointer
(\texttt{0xffffffff82060f20}) appended from \texttt{chain\_config}.
The chain is loaded by \texttt{exploit\_hxp\_generic}, which
constructs the overflow payload, places the gadget sequence
on the stack, and appends the return frame via
\texttt{swapgs\_restore\_regs\_and\_return\_to\_usermode+0x79}.
Execution in the QEMU environment produces \texttt{[+]~GOT~ROOT!}
followed by \texttt{/usr/bin/id} returning \texttt{uid=0, gid=0}.

To further characterize how gadget selection quality affects
exploit outcome, we repeat the experiment with the rank-8 chain, which
contains the \texttt{load\_rdx} gadget
\texttt{pop~rdx;~pop~rdi;~ret} (cost 5). This gadget pops two
values from the stack rather than one, displacing all subsequent
gadget addresses by one slot and corrupting the chain layout.
Execution with this chain produces a kernel panic.
This result illustrates a modeling gap: stack-delta constraints,
rather than gadget cost alone, determine exploit success in this case.

\section{Related Work}
\label{sec:related}

\textbf{ROP automation.}
Gadget chaining as a $W\oplus X$ bypass was established
by~\cite{shacham2007geometry,roemer2012rop}; extensions include
JOP~\cite{bletsch2011jump} and return-into-libc~\cite{checkoway2010return}.
Carlini and Wagner~\cite{carlini2014rop} showed gadget availability
is practically unbounded in large binaries.
Automated tools---ROPgadget~\cite{ropgadget},
angrop~\cite{angrop}, Q~\cite{schwartz2011q},
ROPecker~\cite{cheng2014ropecker},
pwntools~\cite{pwntools}---rely on pattern matching and heuristic
search. AEG systems~\cite{avgerinos2011aeg} use constraint
satisfaction for exploit generation but not binary quadratic
formulations. None formulates ROP gadget selection as a QUBO
suitable for QAOA.

\textbf{Quantum optimization and cybersecurity.}
QAOA~\cite{farhi2014qaoa} has been applied to
MaxCut~\cite{harrigan2021quantum}, portfolio
selection~\cite{egger2021quantum}, and network
routing~\cite{vikstahl2020quantum}.
Guerreschi and Matsuura~\cite{guerreschi2019qaoa} established that
quantum advantage requires hardware improvements beyond current NISQ,
consistent with our findings.
QUBO frameworks and Ising mappings are reviewed
in~\cite{kochenberger2014qubo,glover2019tutorial,lucas2014ising}.
Quantum risk to cybersecurity has been studied primarily through
the lens of cryptography~\cite{shor1994,nist2024pqc,
bernstein2017post,mosca2018cybersecurity}. This work explores
a different direction: applying quantum combinatorial optimization
to offensive security, specifically to ROP gadget selection for
exploit construction.

\section{Discussion and Limitations}
\label{sec:discussion_and_limitations}

\subsection{Scope and Optimality}

This work demonstrates QAOA participating in a real exploit
pipeline. At the problem sizes evaluated here, QAOA is not
faster than greedy or any classical method tested: greedy finds
the retained-set optimum in milliseconds on every instance,
while a single QAOA job requires several seconds of hardware
time plus hours of parameter optimization on a simulator.
GR and ES match the retained-set optimum in all 16 instances
because each pool contains at least one clean gadget per role,
making cross-penalty interactions irrelevant at this scale.
The value of this work lies not in speed but in demonstrating
that a quantum processor can participate in a functional exploit
pipeline and return valid, ranked results.

The problem becomes non-trivial when the pool is dominated by
dirty gadgets: the execve chain on libc.so.6 projects
$\approx860{,}000$ combinations; setresuid with 4 dirty roles
projects $\approx116$ million, where exhaustive search becomes
costly and greedy is expected to diverge from the optimum.

The $N$ lowest-cost gadgets retained per role may exclude a
gadget with slightly higher individual cost but fewer
cross-penalties. In our corpus the cheapest gadgets consistently
have the fewest clobbers, so this approximation is well-aligned
with the objective. All optimality claims refer to the retained
candidate set.

$V_{\mathrm{rate}}$ and $P(\text{optimal})$ measure different
things: the former counts distinct valid outcomes as a fraction
of distinct observed bitstrings; the latter counts optimal shots
over total shots. We report $P(\text{optimal})$ for the hxpCTF
case study only; extending it to all benchmark runs is left as
future work.

\subsection{Modeling Gaps and Reproducibility}

The cost model captures register clobbering but not
execution-context or stack-delta constraints.
\texttt{syscall} from kernel mode is unsuitable as a chain
terminator (it enters \texttt{entry\_SYSCALL\_64}); this was
resolved by choosing \texttt{commit\_creds} as the objective
before gadget extraction.
The dirty gadget \texttt{pop~rdx;~pop~rdi;~ret} passes
semantic verification but consumes an extra stack word,
displacing subsequent addresses. Adding stack delta as an
explicit QUBO penalty is a concrete future improvement.

All QAOA-HW experiments were on IBM Heron r2; the
depth--$V_{\mathrm{rate}}$ correlation will differ on other
backends. The hxpCTF case study uses KASLR-disabled;
in production an address leak is required before applying
the pipeline. The QUBO and circuit are address-agnostic
once the base is known.

\textbf{Future directions:}
(1) instances where greedy fails, demonstrating non-trivial
QUBO structure;
(2) stack-delta in the cost model;
(3) $N\geq5$ on large binaries (execve/setresuid on libc);
(4) warm-starting QAOA from greedy;
(5) evaluation on lower-noise backends.


\section{Conclusion}
\label{sec:conclusion}

We formulated ROP gadget selection as a QUBO and executed it on IBM
Heron r2 as part of a functional Linux kernel exploit pipeline.
A 10-qubit QAOA circuit (depth~185, 80 ECR gates, 4,096 shots)
yields 23 valid chains from the hxpCTF 2020 \texttt{kernel-rop}
challenge; the minimum-cost chain produces
\texttt{uid=0, gid=0} on Linux 5.9.0-rc6+ with SMEP and SMAP
active.

QAOA-HW recovers the retained-set optimum in at least one of
five jobs for 11/16 benchmark instances, with failures
concentrated above transpiled depth~200. A control experiment
confirms the result: a higher-cost chain containing a dirty
gadget produces a kernel panic, while the minimum-cost
QAOA-selected chain achieves privilege escalation. This
establishes that gadget cost, as modelled by the QUBO,
is a reliable signal for exploit success at this scale.

\bibliographystyle{elsarticle-num}
\bibliography{Quantum ROP}

\end{document}